# A Polynomial Diophantine Generator Function for Integer Residuals

Charles Sauerbier

January 2010

*Abstract: Two Diophantine equation generator function for integer residuals produced by integer division over closed intervals are presented. One each for the closed intervals $[1, \lfloor\sqrt[2]{n}\rfloor]$ and $[\lceil\sqrt[2]{n}\rceil, n]$, respectively.*

1. **Preliminaries**

In this paper we address the problem of determining residual values in integer division by mathematical computation more amenable to analytic methods than iterative division, as well as the values of the corresponding integer quotient. Diophantine generator functions derived from empirical observation of the pattern present in integer values as a result of sequential execution of the division operation over each of the respective closed integer intervals is presented that admits the direct computation of corresponding residual and quotient by means of polynomial functions. Unlike the process of simple iterative division the resulting generator functions admit application of mathematical tools to analysis of applicable problems not otherwise admitted by the sequential division process.

We have not found the results presented addressed by others in available literature. The focus of the literature of general relation to the problem and problems it impacts having taken avenues premised on more conventional methods, such as algebraic structures and numeric sieves. The approach taken in derivation of the generator functions is by way of difference expression[1] on the 3-tuple <x, y, r> in context of the commonly known relation $n = (x * y) + r$.

2. **Generator Functions**

The observed behavior of the integer values in the two intervals leads us to produce distinct generator functions for each interval. This dichotomy works to benefit the objective of locating the zeros of the function $\left.\frac{n}{x}\right|_{x=1}^{n}$, as the interval $[\lceil\sqrt[2]{n}\rceil, n]$ contains precisely one value for 'x' such that 'x' divides 'n'.

**2.1. Conditions**

We consider the following conditions: Given a value 'n', 3-tuples <$x_k$, $y_k$, $r_k$> where <$x_0$, $y_0$, $r_0$> = $\langle \lfloor\sqrt[2]{n}\rfloor, \lceil n/\sqrt[2]{n}\rceil, n - (\lfloor\sqrt[2]{n}\rfloor * \lceil n/\sqrt[2]{n}\rceil) \rangle$. The values of $x_0$, $y_0$ have the relation $x_0 \leq y_0$. For our purpose here we define on the values i = $x_0$ and j = $y_0$ the following delta: $\delta_{ij} = \begin{cases} 1 \text{ where } i \neq j \\ 0 \text{ where } i = j \end{cases}$.

---

[1] (Kelley & Peterson, 1990) (Elaydi, 2005) (Goldberg, 1986)

## 2.2. X Decrement $[1, x_0]$

The generator functions here are premised on decrementing 'x' over the interval $\lfloor \sqrt[2]{n} \rfloor \geq x \geq 1$, where 'x' is an integer index value. The generator function is derived from n = (x * y) – r. We obtain the set of difference expressions of *Lemma 2.2.1* for the respective values of x, y, r from that relation.

### Lemma 2.2.1

*Given the conditions as stated, above in 2.1, for all values k in the closed interval $[1, x_0]$ over integers:*

$$x_k = x_{k-1} - 1$$
$$y_k = y_{k-1} + ((r_{k-1} + y_{k-1}) \text{ div } x_k), \text{ and;}$$
$$r_k = (r_{k-1} + y_{k-1}) \text{ mod } x_k.$$

The difference expressions follow by basic algebra from the relations $n = (x * y) - r$, as any amount by which x is reduced needs be account for in the other terms of the equation to maintain the equality relation between the left and right sides. The expressions for computing the resultant values of 'y' and 'r' distribute the subtracted value of 'x' in the context of the equation across those variables in accordance with the algebraic manipulation:

$$n = ((x - 1) * y) + (r + y)$$
$$n = \left((x - 1) * \left(y + ((r + y) \text{ div } (x - 1))\right)\right) + ((r + y) \text{ mod } (x - 1))$$

### Theorem 2.2.1

*Given the conditions as stated, above in 2.1, under the difference expressions of Lemma 2.2.1, where $x_k = x_{k-1} - 1$, the residual value $r_k = r'_k \text{ mod } x_k$ where*

$$r'_k = r_0 + (k * \delta_{ij}) + \left(\frac{(k-1)k}{2}\right) + \left(\frac{((k+1)k)}{2}\right);$$

*which reduces to*

$$r'_k = r_0 + (k * \delta_{ij}) + k^2.$$

We observe that under the given conditions the value $\sqrt[2]{n}$ will be either an integer or real number with a non-zero decimal component. The result of the assignment of values to $x_0, y_0$ is such that either $x_0 = y_0$ or $x_0 = y_0 - 1$ for any given value of 'n'. With $\delta_{ij}$ representing $|y_0 - x_0|$ we have then the follow sequence:

$$r_0 = n - \lfloor x_0 * y_0 \rfloor$$
$$r_1 = r_0 + (y_0 - x_1) = (r_0 + y_0) \text{ mod } x_1$$
$$r_2 = r_1 + (y_1 - x_2) = (r_1 + y_1) \text{ mod } x_2$$

The sequence of relations above only hold where $(y_{i-1} - x_i) = (r_{i-1} + x_{i-1}) \bmod x_i$. However, in foregoing the application of the mod and div operations we arrive at the following:

$$r'_0 = r_0$$
$$r'_1 = r_0 + (y_0 - x'_1),\ x'_1 = x_0 - 1,\ y'_1 = y_0 + 1$$

Which then leads to the following:

$$r'_1 = r_0 + (y_0 - (x_0 - 1)) = r_0 + (y_0 - x_0) + 1$$
$$r'_2 = r'_1 + (y'_1 - x'_2) = r_0 + (y_0 - x_0) + 1 + (y'_1 - x'_2)$$
$$= r_0 + (y_0 - x_0) + 1 + ((y_0 + 1) - (x_0 - 2))$$
$$= r_0 + (y_0 - x_0) + 1 + (y_0 - x_0) + (1 + 2)$$

Observing that $(y_0 - x_0)$ is equal to 0 or 1 and present in each iteration we have for any $r'_k$ the product $(k * \delta_{ij})$. This leaves the summation of $y_{i-1} - x_i$ for each $r'_k$ giving us

$$\sum_{i=1}^{k-1} i + \sum_{i=1}^{k} i.$$

From which we can thus derive:

$$r'_k = r_0 + (k * \delta_{ij}) + \left(\frac{(k-1)k}{2}\right) + \left(\frac{((k+1)k)}{2}\right);$$

which by algebraic operations on the summations reduces to

$$r'_k = r_0 + (k * \delta_{ij}) + k^2.$$

Testing this hypothesis for base values 0 and 1 we obtain the following results:

$$r'_0 = r_0 + (0 * \delta_{ij}) + 0^2 = r_0$$
$$r'_1 = r_0 + (1 * \delta_{ij}) + 1^2 = r_0 + \delta_{ij} + 1.$$

By induction we then assume that the expression holds for all k, for k+1 we then have

$$r'_{k+1} = r_0 + ((k+1) * \delta_{ij}) + (k+1)^2$$
$$r'_{k+1} = r_0 + (k * \delta_{ij}) + \delta_{ij} + (k^2 + 2k + 1)$$
$$r'_{k+1} = r_0 + (k * \delta_{ij}) + k^2 + (2k+1) + \delta_{ij}$$
$$r'_{k+1} = r_k + (2k+1) + \delta_{ij}$$

Observing that absent application of the quotient of $r'_k\ div\ x_k$ that the difference between the corresponding values of $x'_{k+1}, y'_k$ increase with k by 2, including the constant 1 resulting from the decrementing action of $x_k = x_{k-1} - 1$ the above then reduces to

$$r'_{k+1} = r'_k + (y'_k - x'_{k+1}),$$

proving the theorem.

### Theorem 2.2.2

*Given the conditions as stated, above in 2.1, under the difference expressions of Lemma 2.2.1, as applied in Theorem 2.2.1:*

$$ax'_k = r'_k \text{ only where } x'_k \text{ divides } n \text{ for all values of } a.$$

$$by'_k = r'_k \text{ only where } y'_k \text{ divides } n \text{ for all values of } b.$$

Let us assume $x'_k \nmid n$ and $ax'_k = r'_k$; then $n = (x'_k * y'_k) + ax'_k = x'_k(y'_k + a)$ implies $x'_k$ divides 'n', in contradiction of the assumption.

Similarly, assume $y'_k \nmid n$ and $by'_k = r'_k$; then $n = (x'_k * y'_k) + ay'_k = y'_k(x'_k + a)$ implies $y'_k$ divides 'n', in contradiction of the assumption.

### Corollary 2.2.1

*Given the conditions as stated, above in 2.1, under the difference expressions of Lemma 2.2.1, as applied in Theorem 2.2.1:*

$$ax_0 = r'_k + ak \text{ only where } x_k \text{ divides } n \text{ for all values of } a.$$

This corollary follows from Theorem 3 by substitution of $x_k = x_0 - k$; where the equality follows from the definition of $x_k$ as the $k^{th}$ subtraction of the value 1.

### Lemma 2.2.2

*Given the conditions as stated, above in 2.1, under the difference expressions of Lemma 2.2.1, as applied in Theorem 2.2.1:*

$$y'_k = y_0 + k$$

### Lemma 2.2.3

*Given the conditions as stated, above in 2.1, under the difference expressions of Lemma 2.2.1, as applied in Theorem 2.2.1 and using the result of Lemm2.2.2:*

$$y_k = y'_k + r'_k div\ x_k$$

Noting that the value computed as $y'_k$ is absent the addition of integer quotient of $\left\lfloor \frac{r}{x} \right\rfloor$ (i.e. "r div x") and that $r'_k$ accounts for that integer quotient in the absence of the mod operation being applied, from *Lemma 2.2.1* it follows that $y_k = y'_k + r'_k div\ x_k$, producing a generator function for the quotient of integer division of 'n' by 'x'.

## 2.3. Y-Increment $[y_0, n]$

The generator functions here are premised on incrementing 'y' over the interval $\lfloor \sqrt[2]{n} \rfloor \leq y \leq n$, where 'y' is an integer index value. The generator function is derived from n = (x * y) – r. We obtain the set of difference expressions of *Lemma 2.1.1* for the respective values of x, y, r from that relation.

The generator functions presented in this section operate over the interval $\lceil \sqrt[2]{n} \rceil \leq y \leq n$. Computation of 'x' without modular normalization of values will cause 'x' to become negative, which in computer applications may become problematic.

*Lemma 2.3.1*

*Given the conditions as stated, above in 2.1, for all values k in the closed interval $[y_0, n]$ over integers:*

$$y_k = y_{k-1} + 1$$
$$x_k = x_{k-1} + ((r_{k-1} - x_{k-1}) \text{ div } y_k), \text{ and;}$$
$$r_k = (r_{k-1} - x_{k-1}) \text{ mod } y_k.$$

The difference expressions follow by basic algebra from the relations $n = (x * y) - r$, as any amount by which y is increases needs be account for in the other terms of the equation to maintain the equality relation between the left and right sides. The expressions for computing the resultant values of 'x' and 'r' absorb the added value of 'y' in the context of the equation across those variables in accordance with the algebraic manipulation:

$$n = ((y + 1) * x) + (r - x)$$
$$n = \left((y + 1) * \left(x + ((r - x) \text{ div } (y + 1))\right)\right) + ((r - x) \text{ mod } (y + 1))$$

*Theorem 2.3.1*

*Given the conditions as stated, above in 2.1, under the difference expressions of Lemma 2.3.1, where $y_k = y_{k-1} + 1$, for all values k in the closed interval $[y_0, n]$ over integers the residual value $r_k = r'_k \text{ mod } y_k$ where*

$$r'_k = r_0 + (k * \delta_{ij}) + \left(\frac{(k-1)k}{2}\right) + \left(\frac{((k+1)k)}{2}\right);$$

*which reduces to*

$$r'_k = r_0 + (k * \delta_{ij}) + k^2.$$

We observe that under the given conditions the value $\sqrt[2]{n}$ will be either an integer or real number with a non-zero decimal component. The result of the assignment of values to $x_0, y_0$ is such that either $x_0 = y_0$ or $x_0 = y_0 - 1$ for any given value of 'n'. With $\delta_{ij}$ representing $|y_0 - x_0|$ we have then the follow sequence:

$$r_0 = n - \lfloor x_0 * y_0 \rfloor$$
$$r_1 = r_0 + (x_0 - y_1) = (r_0 + x_0) \text{ mod } y_1$$
$$r_2 = r_1 + (x_1 - y_2) = (r_1 + x_1) \text{ mod } y_2$$

The sequence of relations above only hold where $(x_{i-1} - y_i) = (r_{i-1} + x_{i-1}) \bmod y_i$. However, in foregoing the application of the mod and div operations we arrive at the following:

$$r'_0 = r_0$$
$$r'_1 = r_0 + (x_0 - y'_1), \; x'_1 = x_0 - 1, \; y'_1 = y_0 + 1$$

Which then leads to the following:

$$r'_1 = r_0 + (x_0 - (y_0 + 1)) = r_0 + (x_0 - y_0) + 1$$
$$r'_2 = r'_1 + (x'_1 - y'_2) = r_0 + (x_0 - y_0) + 1 + (x'_1 - y'_2)$$
$$= r_0 + (x_0 - y_0) + 1 + ((x_0 - 1) - (y_0 - 2))$$
$$= r_0 + (x_0 - y_0) + 1 + (x_0 - y_0) + (1 + 2)$$

Observing that $(y_0 - x_0)$ is equal to 0 or 1 and present in each iteration we have for any $r'_k$ the product $(k * \delta_{ij})$. This leaves the summation of $x_{i-1} - y_i$ for each $r'_k$ giving us

$$\sum_{i=1}^{k-1} i + \sum_{i=1}^{k} i.$$

From which we can thus derive:

$$r'_k = r_0 + (k * \delta_{ij}) + \left(\frac{(k-1)k}{2}\right) + \left(\frac{((k+1)k)}{2}\right);$$

which by algebraic operations on the summations reduces to

$$r'_k = r_0 + (k * \delta_{ij}) + k^2.$$

The proof then follows in consequence of Theorem 2.2.1.

### Theorem 2.3.2

*Given the conditions as stated, above in 2.1, under the difference expressions of Lemma 2.3.1, as applied in Theorem 2.3.1, for all values k in the closed interval $[y_0, n]$ over integers:*

$$ax'_k = r'_k \; only \; where \; x'_k \; divides \; n \; for \; all \; values \; of \; a.$$
$$by'_k = r'_k \; only \; where \; y'_k \; divides \; n \; for \; all \; values \; of \; b.$$

Let us assume $x'_k \nmid n$ and $ax'_k = r'_k$; then $n = (x'_k * y'_k) + ax'_k = x'_k(y'_k + a)$ implies $x'_k$ divides 'n', in contradiction of the assumption.

Similarly, assume $y'_k \nmid n$ and $by'_k = r'_k$; then $n = (x'_k * y'_k) + ay'_k = y'_k(x'_k + a)$ implies $y'_k$ divides 'n', in contradiction of the assumption.

### Corollary 2.3.1

*Given the foregoing conditions under the difference expressions of Lemma 2.3.1, as applied in Theorem 2.3.1, for all values k in the closed interval $[y_0, n]$ over integers:*

$$ay_0 = r'_k - ak \; only \; where \; y'_k \; divides \; n \; for \; all \; values \; of \; a.$$

This corollary follows from Theorem 3 by substitution of $y'_k = y_0 + k$; where the equality follows from the definition of $y_k$ as the k$^{th}$ addition of the value 1.

*Lemma 2.3.2*

*Given the conditions as stated, above in 2.1, under the difference expressions of Lemma 2.3.1, as applied in Theorem 2.3.1, for all values k in the closed interval $[y_0, n]$ over integers:*

$$x'_k = x_0 + k$$

Follows directly from definition of difference expressions, with the exception that where $k \geq x_0$, the relation does not hold.

*Lemma 2.3.3*

*Given the conditions as stated, above in 2.1, under the difference expressions of Lemma 2.3.1, as applied in Theorem 2.3.1 and using the result of Lemm2.3.2, for all values k in the closed interval $[y_0, n]$ over integers:*

$$x_k = x'_k + r'_k \, div \, y_k$$

Noting that the value computed as $x'_k$ is absent the addition of integer quotient of $\left\lfloor \frac{r}{y} \right\rfloor$ (i.e. "r div y") and that $r'_k$ accounts for that integer quotient in the absence of the mod operation being applied, from *Lemma 2.3.1* it follows that $x_k = x'_k + r'_k \, div \, y_k$, producing a generator function for the quotient of the integer division of 'n' by 'y'.

3. Conclusion

As a result of observation of empirical data sets a Diophantine generator function of integer quotient and residual in integer division was derived. The generator functions admits the computation of both quotient and residual values using continuous polynomial functions on basis of an index in either of the closed integer intervals $[1, \lfloor \sqrt[2]{n} \rfloor]$ or $[\lceil \sqrt[2]{n} \rceil, n]$. The generator functions provide means to determine the zeros within either closed interval where 'n' is the dividend. The results presented have potential application in determining the factors of an integer value, as well as implications for other applications.

4. References

Elaydi, S. (2005). *An Introduction to Difference Equations.* Springer.

Goldberg, S. (1986). *Introduction to Difference Equations.* Dover Publications.

Kelley, W. G., & Peterson, A. C. (1990). *Difference Equations : An introduction with Applications.* Academic Press.